\documentclass[runningheads]{svmult}

\usepackage{makeidx}   
\usepackage{graphicx}  
\usepackage{subeqnar}  
\usepackage{multicol}  
\usepackage{physprbb}  
\makeindex             

\begin{document}
\title*{The Most Luminous Galaxies}
\toctitle{Luminous Galaxies}
%
%
\titlerunning{Merger Driven Galaxies}
%
\author{I. F\'elix Mirabel\inst{1,2}}
\authorrunning{I. F\'elix Mirabel}
%
%
\institute{CEA/DSM/DAPNIA, Service d'Astrophysique, 91191 Gif/Yvette.
  France \and IAFE/CONICET. cc 67, suc 28. Ciudad Universitaria. 1428 Buenos
  Aires. Argentina}

\maketitle              

\begin{abstract}
  Ultraluminous galaxies in the local universe (z$\leq$0.2) emit the
  bulk of their energy in the mid and far-infrared. The multiwavelength 
approach to these objects has shown that they are advanced mergers 
of gas-rich spiral galaxies. Galaxy-galaxy
  collisions took place on all cosmological time-scales, and nearby mergers serve as
  local analogs to gain insight into the physical processes that lead
  to the formation and trans-formation of galaxies in the more distant
  universe. Here I review multiwavelength observations --with
  particular emphasis on recent results obtained with ISO-- of mergers
  of massive galaxies driving the formation of: 1) luminous infrared
  galaxies, 2) elliptical galaxy cores, 3) luminous dust-enshrouded
  extranuclear starbursts, 4) symbiotic galaxies that host AGNs, and 5) tidal dwarf
  galaxies. The most important implication for studies on the
  formation of galaxies at early cosmological timescales is that the 
distant analogs to the local ultraluminous infrared  galaxies 
are invisible in the
  ultraviolet and optical wavelength rest-frames and should be detected 
as sub-millimeter sources with no optical counterparts.
\end{abstract}

\section{Luminous Galaxies}

One of the most important discoveries from extragalactic observations
at mid- and far-infrared wavelengths has been the identification of a
class of ``Luminous Infrared Galaxies'' (LIGs), objects that emit more
energy in the infrared ($\sim${\ts}5--500{\ts}$\mu m$) than at all
other wavelengths combined (see \cite{SandersMirabel} for a
comprhensive review).  The first all-sky survey at far-infrared
wavelengths carried out in 1983 by the {\it Infrared Astronomical
  Satellite} ({\it IRAS}) resulted in the detection of tens of
thousands of galaxies, the vast majority of which were too faint to
have been included in previous optical catalogs.  It is now clear that
part of the reason for the large number of detections is the fact that
the majority of the most luminous galaxies in the Universe are
extremely dusty.  Previous assumptions, based primarily on optical
observations, about the relative distributions of different types of
luminous galaxies---e.g.  starbursts, Seyferts, and quasi-stellar
objects (QSOs)---need to be revised.

Galaxies bolometrically more luminous than $\sim${\ts}4{\ts}${\it
  L}^*$ (i.e. {\it L}$_{bol} \geq 10^{11}{\ts}{\it L}_\odot$) appear
to be heavily obscured by dust.  Although luminous infrared galaxies
(hearafter LIGs: {\it L}$_{ir} >{\ts}10^{11}{\ts}{\it L}_\odot$) are
relatively rare objects, reasonable assumptions about the lifetime of
the infrared phase suggest that a substantial fraction of all galaxies
with {\it L}$_B$ $>{\ts}10^{10}${\it}{\it L}$_\odot$ pass through such
a stage of intense infrared emission \cite{Soifer1}.

\begin{figure}[htb]
\begin{center}
\includegraphics[width=.8\textwidth]{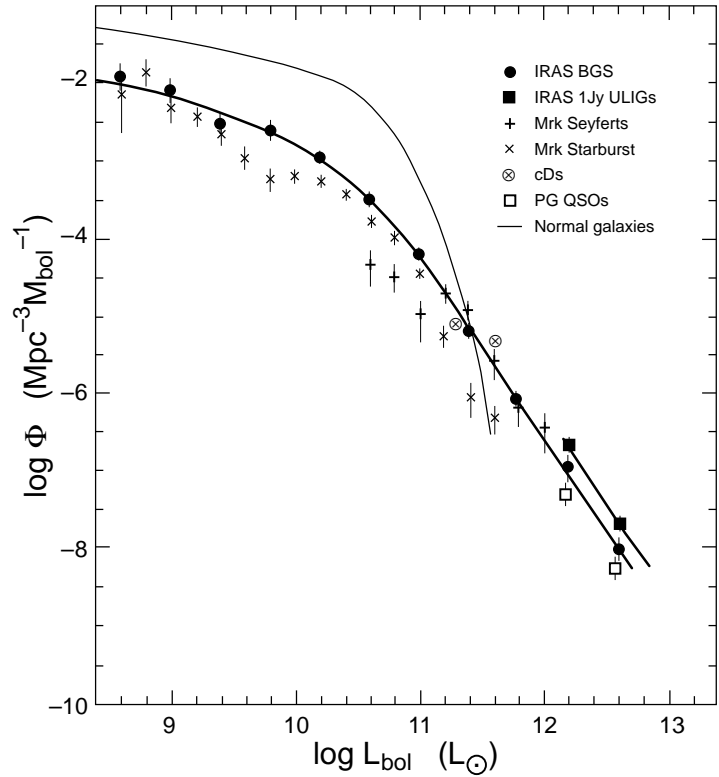}
\end{center}
\caption[]{Galaxy luminosity function of Infrared Galaxies compared
  with other extragalactic objects in the local universe. Among the
  most luminous galaxies (${\it L}_{bol} >
  10^{11.5}${\ts}{\it}L$_\odot$), infrared galaxies selected from the
  IRAS survey outnumber optically selected Seyferts and quasars.  For
  references see \cite{SandersMirabel}.}
\label{fig1}
\end{figure}

A comparison of the luminosity function of infrared bright galaxies
with other classes of extragalactic objects in the local universe is
shown in Figure \ref{fig1}. At luminosities below $10^{11}${\ts}{\it
  L}$_\odot$, {\it IRAS} observations confirm that the majority of
optically selected objects are relatively weak far-infrared emitters.
Surveys of Markarian galaxies confirm that both Markarian starbursts
and Seyferts have properties (e.g. ${\it f}_{60}/{\it f}_{100}$ and
${\it L}_{ir} / {\it L}_B$\ ratios) closer to infrared selected
samples as does the subclass of optically selected interacting
galaxies. However because the most luminous galaxies are enshrouded in
dust, relatively few objects in optically selected samples are found
with ${\it L}_{ir} > 10^{11.5}${\ts}{\it}L$_\odot$.

\begin{figure}[htb]
\begin{center}
\includegraphics[width=.8\textwidth]{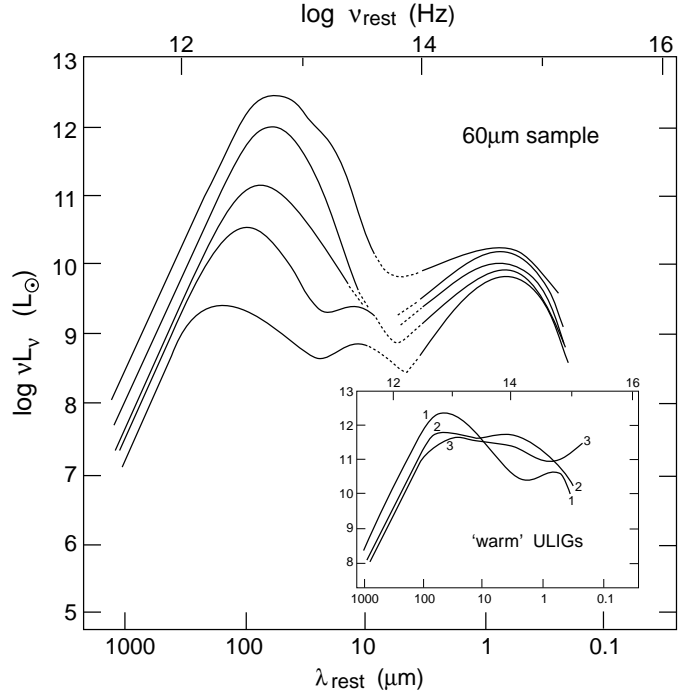}
\end{center}
\caption[]{ Variation of the mean Spectral Energy Distribution (from
  submillimeter to UV wavelengths) with increasing { L}$_{ir}$ for a
  60{\ts}$\mu$m sample of infrared galaxies.  ({ Insert}) Examples of
  the subset ($\sim${\ts}15\%) of ULIGs with ``warm'' infrared color
  ({ f}$_{25}$/{ f}$_{60}$ $>${\ts}0.3).  Three objects (1---the
  powerful Wolf-Rayet galaxy IRAS{\ts}01002--2238, 2---the ``infrared
  QSO'' IRAS{\ts}07598+6508, 3---the optically selected QSO
  I{\ts}Zw{\ts}1) are shown in the inset. For references see
  \cite{SandersMirabel}.}
\label{fig2}
\end{figure}

The high luminosity tail of the infrared galaxy luminosity function is
clearly in excess of what is expected from the Schechter function.
For ${\it L}_{bol} = 10^{11}-10^{12}${\ts}{\it L}$_\odot$, LIGs are as
numerous as Markarian Seyferts and $\sim${\ts}3 times more numerous
than Markarian starbursts.  Ultraluminous infrared galaxies (hereafter
ULIGs: ${\it L}_{ir} > 10^{12}${\ts}{\it L}$_\odot$) appear to be
$\sim${\ts}2 times more numerous than optically selected QSOs, the
only other previously known population of objects with comparable
bolometric luminosities.

\begin{figure}[htb]
\begin{center}
\includegraphics[width=.8\textwidth]{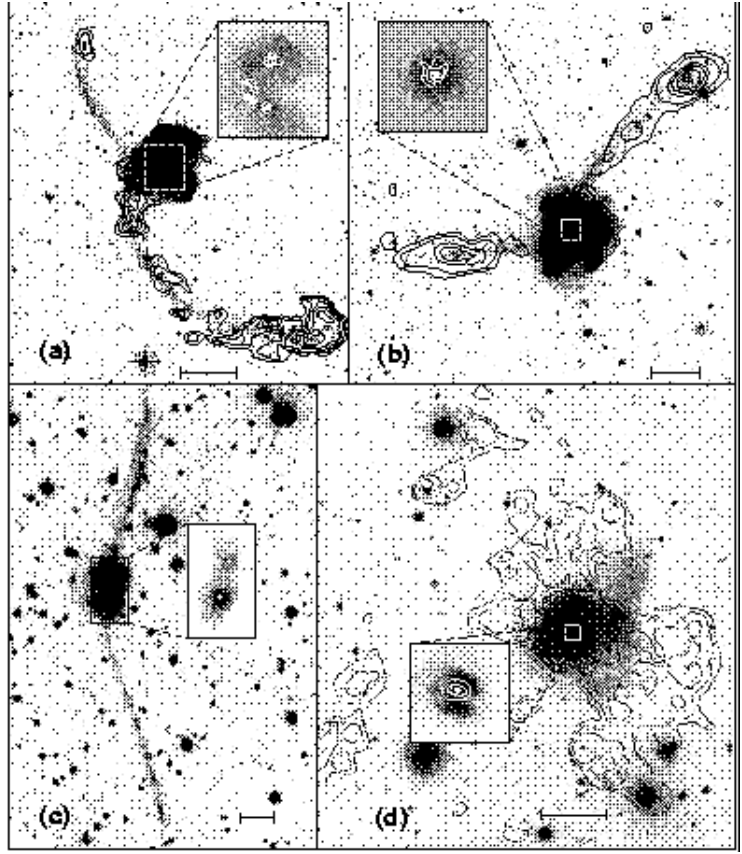}
\end{center}
\caption[]{ Well-studied mergers: 
  {\it (a)}\ NGC{\ts}4038/39 (Arp{\ts}244 = ``The Antennae''); {\it
    (b)}\ NGC{\ts}7252 (Arp{\ts}226 = ``Atoms for Peace''); {\it (c)}\ 
  IRAS{\ts}19254--7245 (``The Super Antennae''); {\it (d)}\ 
  IC{\ts}4553/54 (Arp{\ts}220).  The two at the top are LIGs whereas
  the two at the bottom are ULIGs.  Contours of H{\ts}I 21-cm line
  column density ({\it black}) are superimposed on deep optical ({\it
    r}-band) images.  Inserts show a more detailed view in the {\it
    K}-band (2.2{\ts}$\mu$m) of the nuclear regions of
  NGC{\ts}4038/39, NGC{\ts}7252, and IRAS{\ts}19254--7245, and in the
  {\it r}-band (0.65{\ts}$\mu$m) of Arp{\ts}220.  White contours
  represent the CO(1$\to$0) line integrated intensity as measured by
  the OVRO millimeter-wave interferometer.  No H{\ts}I or CO
  interferometer data are available for the southern hemisphere object
  IRAS{\ts}19254--7245. The scale bar represents 10{\ts}kpc.}
\label{fig3}
\end{figure}

Although LIGs comprise the dominant population of extragalactic
objects at ${\it L}_{bol} > 10^{11}${\ts}{\it L}$_\odot$, they are
still relatively rare.  For example, Figure \ref{fig1} suggests that
only one object with ${\it L}_{ir} >{\ts}10^{12}${\ts}{\it L}$_\odot$\ 
will be found out to a redshift of $\sim${\ts}0.033, and indeed,
Arp{\ts}220 (${\it z} = 0.018$) is the only ULIG within this volume.
The total infrared luminosity from LIGs in the {\it IRAS} Bright
Galaxy Survey (BGS) is only $\sim${\ts}6\% of the infrared emission in
the local Universe \cite{Soifer2}.

There are preliminary indications that ULIGs have been more numerous
in the past.  Comparison of the space density of nearby ULIGs with the
more distant population provides evidence for possible strong
evolution in the luminosity function at the highest infrared
luminosities.  Assuming pure density evolution of the form $\Phi ({\it
  z}) \propto (1 + {\it z})^{ n}$, \cite{Kim} found ${\it n}
\sim{\ts}7\pm${\ts}3 for a complete flux-limited sample of ULIGs.

\begin{figure}[p]
\begin{center}
\includegraphics[width=.8\textwidth]{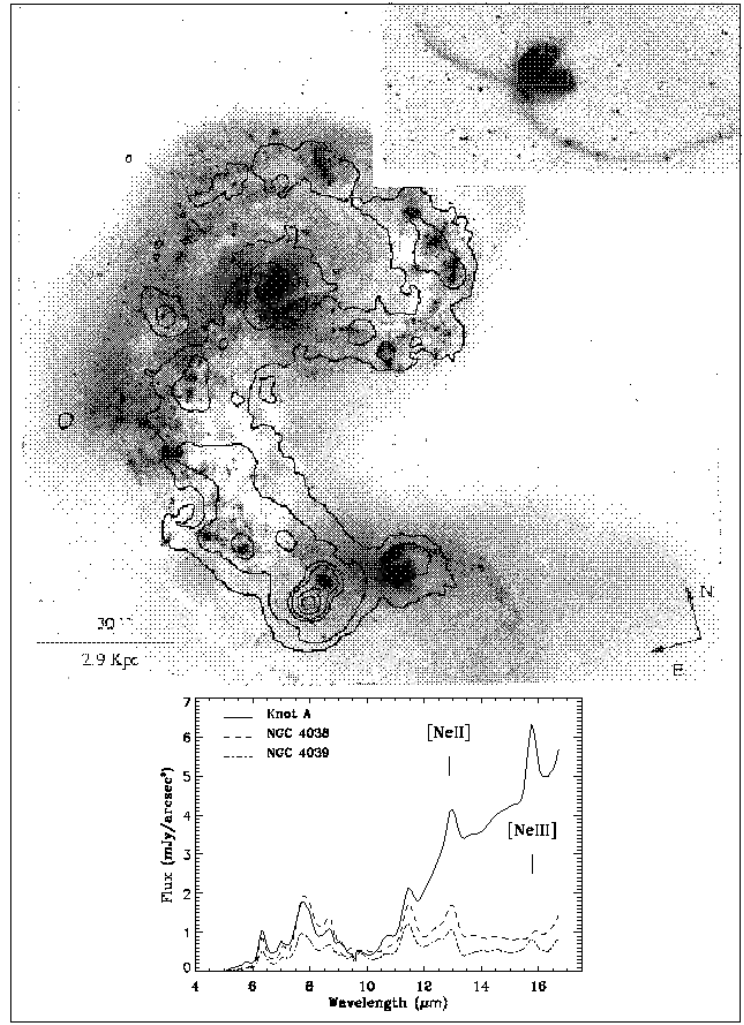}
\end{center}
\caption[]{The upper figure from \cite{Mirabel98} shows a superposition of
  the mid-infrared (12 -17 $\mu$, countours) image of the Antennae
  galaxies obtained with the Infrared Space Observatory, on the
  composite optical image with V (5252 \AA) and I (8269 \AA) filters
  recovered from the Hubble Space Telescope archive .  About half of
  the mid-infrared emission from the gas and dust that is being heated
  by recently formed massive stars comes from an off-nuclear region
  that is clearly displaced from the most prominent dark lanes seen in
  the optical. The brightest mid-infrared emission comes from a region
  that is relativelly inconspicuous at optical wavelengths. The ISOCAM
  image was made with a 1.5$''$ pixel field of view. Contours are 0.4,
  1, 3, 5, 10, and 15 mJy.  The lower figure shows the spectrum of the
  brightest mid-infrared knot and of the nuclei of NGC 4038 and NGC
  4039. The rise of the continuum above 10 $\mu$m and strong NeIII
  line emission observed in the brightest mid-infrared knot indicate
  that the most massive stars in this system of interacting galaxies
  are being formed in that optically obscured region, still enshrouded
  in large quantities of gas and dust.}
\label{fig4}
\end{figure}

The infrared properties for the complete {\it IRAS} Bright Galaxy
Sample have been summarized and combined with optical data to
determine the relative luminosity output from galaxies in the local
Universe at wavelengths $\sim${\ts}0.1--1000{\ts}$\mu m$
\cite{Soifer2}.  Figure \ref{fig2} illustrates how the shape of the
mean spectral energy distribution (SED) varies for galaxies with
increasing total infrared luminosity.  Systematic variations are
observed in the mean infrared colors; the ratio ${\it f}_{60}/{\it
  f}_{100}$ increases while ${\it f}_{12}/{\it f}_{25}$ decreases with
increasing infrared luminosity. Figure \ref{fig2} also illustrates
that the observed range of over 3 orders of magnitude in ${\it
  L}_{ir}$\ for infrared-selected galaxies is accompanied by less than
a factor of 3--4 change in the optical luminosity.

\cite{Sandersb} showed that a small but significant fraction of ULIGs,
those with ``warm'' ({\it f}$_{25}$/{\it f}$_{60}$ $>${\ts}0.3)
infrared colors, have SEDs with mid-infrared emission
($\sim$5--40{\ts}$\mu$m) over an order of magnitude stronger than the
larger fraction of ``cooler'' ULIGs.  These warm galaxies (Figure
\ref{fig2} insert), which appear to span a wide range of classes of
extragalactic objects including powerful radio galaxies (PRGs: ${\it
  L}_{408MHz} \geq 10^{25} W Hz^{-1}$) and optically selected QSOs,
have been used as evidence for an evolutionary connection between
ULIGs and QSOs (e.g. \cite{Sandersa,Sandersb}).

There is a strong correlation between the broad band colors (from
optical to far-infrared) and morphological type \cite{SandersMirabel}.
In particular, the fraction of objects that are interacting/merger
systems appears to increase systematically with increasing infrared
luminosity.  The imaging surveys of objects in the local universe
\cite{Sandersa,Melnick} have shown that the fraction of strongly
interacting/merger systems increases from $\sim${\ts}10\% at $log\ 
({\it L}_{ir} / {\it L}_{\odot}$) = 10.5--11 to $\sim${\ts}100\% at
$log\ ({\it L}_{ir} / {\it L}_{\odot}$) $>${\ts}12. In pannel (c) of
Figure \ref{fig3} is shown the "Super-antennae", which is the
prototype of ULIG \cite{Mirabel91}. ISO observations \cite{Laurent99}
have shown that more than 98\% of the mid-infrared flux from this
object comes from the southern component which hosts a Seyfert 2
nucleus.

From the detailed studies of nearby ultraluminous infrared galaxies
the following conclusions were reached.  1) They are mergers of
evolved gas-rich giant spiral galaxies (e.g. Milky Way with
Andromeda), and not ``primival" galaxies.  2) To boost the luminosity
above 10$^{12}$ L$_{\odot}$ the nuclei must have approached at least
10 kpc, namely, they are advanced mergers.  3) Due to the
gravitational impact the interstellar gas decouples from the stars and
large amounts of interstellar matter fall at high rates to the central
region.  This is the condition to produce a nuclear starburst, and/or
feed a supermassive black hole at super-Eddington accretion rates. To
produce such large accretion rates, the gravitational potential wheels
of massive buldges are needed.

A workshop on the question concerning the ultimate source of energy
(starbursts versus AGN's) took place in Ringberg on October 1998.
Below 2 10$^{12}$ L$_{\odot}$ starbursts dominate the energy budget,
but above 3 10$^{12}$ L$_{\odot}$ AGN's seem to be always present and
become an important source of energy. In this respect it is
interesting to note that it is found with ISO that in the prototype
Seyfert 2 galaxy NGC 1068, about 80\% of the mid-infrared flux between
4 and 18 $\mu$m comes from the AGN \cite{LeFloc}.

A caveat for the subject of this conference is that the pre-encounter
objects that merged at high redshifts must have been different from
the metal-rich evolved galaxies merging at present. Another caveat is
that ultraluminous IR galaxies at high redshifts may be very difficult
to detect using the Lyman break technique. Due to the large amounts of
dust in ultraluminous objects, very little or none continuum leaks out
at ultraviolet wavelengths. Therefore, surveys with submillimeter
arrays as ALMA will be needed to detect ultraluminous galaxies at high
redshifts.

\section{ISO Observation of Extranuclear Starbursts}

The starbursts in ultraluminous galaxies take place in the nuclear
region.  One of the new findings with ISO is a class of very luminous
dust-enshrouded extranuclear starbursts in nearby spiral-spiral
mergers.  When the pre-encounter galaxies do not have prominent
buldges, namely, when the mergers are -for instance- two Sc galaxies,
the most luminous starbursts may take place in extranuclear regions
that are inconspicuous at optical wavelengths.  These extranuclear
starbursts have sizes $\leq$ 100 pc in radius and can produce up to
50\% of the overall mid-infrared output from these systems.
Furthermore, the analyses of the mid-infrared spectra indicate that
the most massive stars in these systems are formed inside these
optically invisible knots.

In Figure \ref{fig4} is shown in contours the mid-infrared (12-17
$\mu$m) image of the Antennae galaxies obtained with ISO
\cite{Mirabel98}, superimposed on the optical image from HST. Below
are shown representative spectra of the two nuclei and the brightest
mid-infrared knot. It shows that the most massive stars are formed in
an obscured knot of 50 pc radius, which produces about 15\% of the
total luminosity from the Antennae galaxies between 12.5 and 17
$\mu$m. A more extreme case is found in NGC 3690 \cite{Gallais}, where
it is observed an extranuclear region $\leq$100 pc in radius that
radates $\sim$45\% of the overall mid-infrared output from this
system. If the fraction of far-infrared fluxes is the same as in the
mid-infrared, such compact region produces a luminosity of 2 10$^{11}$
L$_{\odot}$.  Therefore, the luminosity of a few compact starburst
knots of this type would be comparable to the total bolometric
luminosity of a ULIG such as Arp 220 (Figure \ref{fig3}d).

The multiwavelength view of this nearby sample of prototype merging
systems suggests caution in deriving scenarios of early evolution of
galaxies at high redshift using only observations in the narrow
rest-frame ultraviolet wavelength range \cite{Mirabel98}.  Although
the actual numbers of this type of systems may not be large, we must
keep in mind that the most intense starbursts are enshrouded in dust
and no ultraviolet light leaks out from these regions.

\section{Symbiotic Galaxies}

\begin{figure}[htb]
\begin{center}
\includegraphics[width=.8\textwidth]{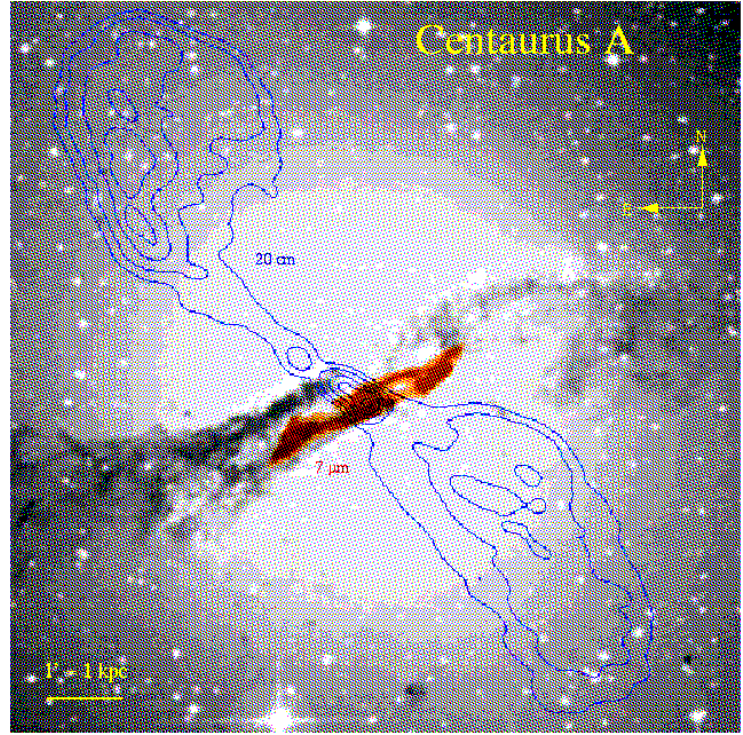}
\end{center}
\caption[]{The ISO 7\,$\mu$m emission (dark structure; \cite{Mirabel99}) and
  VLA 20 cm continuum in contours \cite{Condon}, overlaid on an
  optical image from the Palomar Digital Sky Survey. The emission from
  dust with a bisymmetric morphology at the centre is about 10 times
  smaller than the overall size of the shell structure in the
  elliptical and lies on a plane that is almost parallel to the minor
  axis of its giant host. Whereas the gas associated to the spiral
  rotates with a maximum radial velocity of 250 km s$^{-1}$, the
  ellipsoidal stellar component rotates slowly approximately
  perpendicular to the dust lane \cite{Wilkinson}. The synchrotron
  radio jets shown in this figure correspond to the inner structure of
  a double lobe radio source that extends up to 5$^{\circ}$ ($\sim$
  300 kpc) on the sky. The jets are believed to be powered by a
  massive black hole located at the common dynamic center of the
  elliptical and spiral structures.}
\label{fig5}
\end{figure}

Giant radio galaxies are thought to be massive ellipticals powered by
accretion of interstellar matter onto a supermassive black hole.
Interactions with gas rich galaxies may provide the interstellar
matter to feed the active galactic nucleus (AGN). To power radio lobes
that extend up to distances of hundreds of kiloparsecs, gas has to be
funneled from kiloparsec size scales down to the AGN at rates of
$\sim$1 M$_{\odot}$ yr$^{-1}$ during $\geq$10$^8$ years. Therefore,
large and massive quasi-stable structures of gas and dust should exist
in the deep interior of the giant elliptical hosts of double lobe
radio galaxies. Recent mid-infrared observations with ISO revealed for
the first time a bisymmetric spiral structure with the dimensions of a
small galaxy at the centre of Centaurus A \cite{Mirabel99}. The spiral
was formed out of the tidal debris of accreted gas-rich object(s) and
has a dust morphology that is remarkably similar to that found in
barred spiral galaxies (see Figure \ref{fig5}). The observations of
the closest AGN to Earth suggest that the dusty hosts of giant radio
galaxies like CenA, are ``symbiotic" galaxies composed of a barred
spiral inside an elliptical, where the bar serves to funnel gas toward
the AGN.

The barred spiral at the centre of CenA has dimensions comparable to
that of the small Local Group galaxy Messier 33. It lies on a plane
that is almost parallel to the minor axis of the giant elliptical.
Whereas the spiral rotates with maximum radial velocities of
$\sim$\,250 km s$^{-1}$, the ellipsoidal stellar component seems to
rotate slowly (maximum line-of-sight velocity is $\sim$\,40 km
s$^{-1}$) approximately perpendicular to the dust lane. The genesis,
morphology, and dynamics of the spiral formed at the centre of CenA
are determined by the gravitational potential of the elliptical, much
as a usual spiral with its dark matter halo. On the other hand, the
AGN that powers the radio jets is fed by gas funneled to the center
via the bar structure of the spiral. The spatial co-existence and
intimate association between these two distinct and dissimilar systems
suggest that Cen A is the result from a cosmic symbiosis.

\section{Formation of Ellipticals}

In disk-disk collisions of galaxies, dynamical friction and subsequent
relaxation may produce a mass distribution similar to that in classic
elliptical galaxies.  From the relative numbers of mergers and
ellipticals in the New General Catalogue \cite{Toomre} estimated that
a large fraction of ellipticals could be formed via merging. The first
direct observational evidence for the transition from a disk-disk
merger toward an elliptical was presented in the optical study of
NGC{\ts}7252 by \cite{Schweizer82}. The brightness distribution over
most of the main body of this galaxy which is shown in Figure
\ref{fig3}c is closely approximated by a de Vacouleurs (${\it
  r}^{-1/4}$) profile.  However, NGC{\ts}7252 still contains large
amounts of interstellar gas and exhibits a pair of prominent tidal
tails (see Figure \ref{fig3}b); neither property is typical of
ellipticals.

Near-infrared images are less affected by dust extinction and also
provide a better probe of the older stellar population, which contains
most of the disk mass and therefore determines the gravitational
potential. {\it K}-band images of six mergers by \cite{Wright} showed
that the infrared radial brightness profiles for two
LIGs---Arp{\ts}220 and NGC{\ts}2623---follow an {\it r}$^{-1/4}$ law
over most of the observable disks. Among eight merger remnants,
\cite{Stanford} found {\it K}-band brightness profiles for four
objects that were well fitted by an {\it r}$^{-1/4}$ law over most of
the observable disks.  \cite{Kim} finds a similar proportion
($\sim${\ts}50\%) of ULIGs whose {\it K}-band profiles are well fit by
a {\it r}$^{-1/4}$ law.

More recently, \cite{Kormendy} have proposed that ULIGs are elliptical
galaxies forming by merger-induced dissipative collapse.  The
extremely large central gas densities
($\sim${\ts}$10^2$--$10^3{\ts}{\it M}_{\odot}\ pc^{-3}$) observed in
many nearby ULIGs, and the large stellar velocity dispersions found in
the nuclei of Arp{\ts}220 and NGC{\ts}6240 are comparable to the
stellar densities and velocity dispersions respectively, in the
central compact cores of ellipticals.

Despite the {\it K}-band and CO evidence that LIGs may be forming
ellipticals, we still need to account for two important additional
properties of ellipticals:\ 1.\ the large population of globular
clusters in the extended halos of elliptical galaxies, which cannot be
accounted for by the sum of globulars in two preexisting spirals, and
\ 2.\ the need to remove the large amounts of cold gas and dust
present in infrared-luminous mergers in order to approximate the
relative gas-poor properties of ellipticals. These two issues have
been discussed by \cite{SandersMirabel}.

\begin{figure}[htb]
\begin{center}
\includegraphics[width=.8\textwidth]{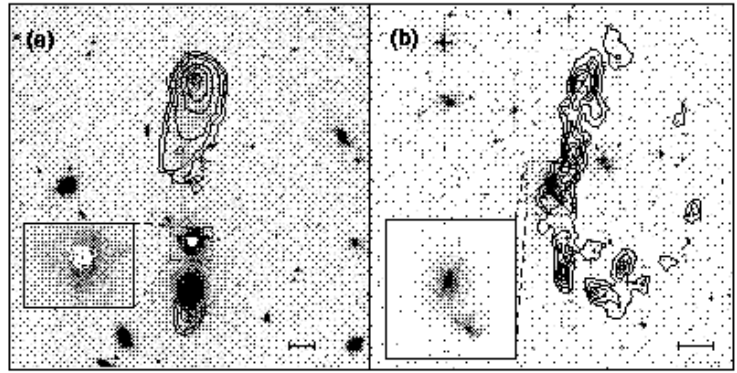}
\end{center}
\caption[]{\it (a)\ NGC{\ts}3561A/B (Arp{\ts}105) from \cite{Duc94};
  \ \ {\it (b)}\ NGC{\ts}5291A/B (``Sea shell'') from \cite{Duc98}.
  Tidal dwarfs may have different morphologies: Blue compacts,
  Magellanic Irregulars, and Dwarf Irregulars. Contours of H{\ts}I
  21-cm line column density ({\it black}) are superimposed on deep
  optical ({\it r}-band) images.  Inserts show a more detailed view in
  {\it r}-band of the spiral galaxy NGC{\ts}3561A \cite{Duc94}, and of
  the interacting pair NGC{\ts}5291A/B.  White contours represent the
  CO(1$\to$0) line integrated intensity as measured by the IRAM
  millimeter-wave interferometer.  CO emission has not been detected
  in NGC{\ts}5291A/B.  The scale bar represents 20{\ts}kpc.}
\label{fig6}
\end{figure}

\section{Tidal Dwarf Galaxies}

Collisions between giant disk galaxies may trigger the formation of
dwarf galaxies. This idea, which was first proposed by \cite{Zwicky}
and later by \cite{Schweizer78}, has received recent observational
support \cite{Mirabel91,Mirabel92,Elmegreen,Duc94}.  Renewed interest
in this phenomenon arose from the inspection of the optical images of
ULIGs, which frequently exhibit patches of optically emitting material
along the tidal tails (see Figure \ref{fig3}a--c).  These objects
appear to become bluer near the tips of the tails at the position of
massive clouds of H{\ts}I. These condensations have a wide range of
absolute magnitudes, {\it M}$_V$ $\sim${\ts}$-14$ to $-19.2$, and
H{\ts}I masses, ${\it M}(H{\ts}I) \sim{\ts}5 \times 10^8$ to $6 \times
10^9${\ts}{\it M}$_\odot$.  \cite{Mirabel95} have shown that objects
resembling irregular dwarfs, blue compacts, and irregulars of
Magellanic type are formed in the tails.  These small galaxies of
tidal origin are likely to become detached systems, namely, isolated
dwarf galaxies. Because the matter out of which they are formed has
been removed from the outer parts of giant disk galaxies, the tidal
dwarfs we observe forming today have a metallicity of about one third
solar \cite{Duc95}.

It is interesting that in these recycled galaxies of tidal origin
there is---as in globular clusters---no compelling evidence for dark
matter \cite{Mirabel95}. To find the true fraction of dwarf
galaxies that may have been formed by processes similar to the tidal
interactions we observe today between giant spiral galaxies, more
extensive observations of interacting systems will be needed. A recent
step forward is the statistical finding that perhaps as much as one
half of the dwarf population in groups is the product of interactions
among the parent galaxies \cite{Hunsberger}.

Tidal dwarfs are formed not only during spiral-spiral mergers, but
also in encounters of spirals with massive ellipticals in clusters of
galaxies.  In Figure \ref{fig6} are shown the results from the
multiwavelength study of Arp 105 and NGC 5291A/B which are in clusters
of galaxies. In Arp 105, \cite{Duc94} find tidal dwarfs that
resemble Magellanic Irregulars and a blue compact. In NGC 5291, about
10 tidal dwarfs of irregular morphology are found associated to the
200 kpc HI ring shown in Figure \ref{fig6} \cite{Duc98}.

\section{Conclusions}

  1) Scenarios on the history of star formation that use only
observations in the UV and optical rest-frames result in luminosity
functions that are strongly biased in the high luminosity end.

2) The most luminous nuclear and off-nuclear starbursts are enshrouded
in dust.  In merging galaxies ISO revealed off-nuclear starburst knots
with sizes $\leq$100 pc that produce bolometric luminosities of up to
2 10$^{11}$ L$_{\odot}$ (e.g. NGC 3690).  A few of these starburst
knots can produce the overall bolometric luminosity of an
ultraluminous galaxy such as Arp 220.

3) The observation with ISO of the nearest AGN to Earth (Centaurus A)
opens the general question on whether the hosts of giant radio
galaxies are symbiotic galaxies composed of spirals at the centre of
giant ellipticals.

4) Mergers of disks can produce metal-rich elliptical galaxy cores.

5) Collisions between giant disk galaxies trigger the formation of
dwarf galaxies out tidal debris. A fraction of these re-cycled
galaxies become detached systems with diverse morphologies: blue
compact dwarfs, dwarf irregulars, and irregulars of Magellanic type.

\vskip .1in
{\it Acknowledgements:} Most of the work review here was carried out 
in collaboration with  D.B. Sanders, P-A. Duc,  V. Charmandaris and 
O. Laurent.

%

\end{document}